\documentclass{llncs}

\usepackage{tabularx}
\newcolumntype{R}{>{\raggedleft\arraybackslash}X}

\begin{document}

\title{\bf Lossless data compression on GPGPU architectures}

\author{Axel Eirola}

\institute{Aalto University School of Science}

\maketitle 
\thispagestyle{empty}

\begin{abstract}
Modern graphics processors provide exceptional computational power, but only for certain computational models. While they have revolutionized computation in many fields, compression has been largely unnaffected. This paper aims to explain the current issues and possibilities in GPGPU compression. This is done by a high level overview of the GPGPU computational model in the context of compression algorithms; along with a more in-depth analysis of how one would implement bzip2 on a GPGPU architecture.

\end{abstract}

\section{Introduction}
Ever since PCs became common household items, and computer gaming hit mainstream attention during the 90s, the graphics processing unit (GPU) industry has been growing steadily in along with the CPU industry. The need for faster and ever more visually stunning graphics has lead a head-on battle between nVidia and AMD for more powerful GPUs, and the consumers money. The colossal computational power on the GPU chips has until lately only been utilized to their full capacity in 3D gaming, but lately with the advent of new architectures is it possible to exploit the computational resources for more general purposes.

General-purpose computing on graphics processing units (GPGPU) gives software designers the massive parallel computational power and memory bandwidth of modern GPUs, at the cost of more limited programming models. Some tasks, such as scientific computing and audio/video processing, are able to utilize these resources without large complications due to their paralellizable nature. Other tasks such as data compression, especially lossless data compression, are not as lucky and have more difficulties exploiting the power of GPUs. This paper focuses on lossless data compression and attempts to build a picture of the incompatibilities between GPGPU computation and data compression. On the other hand it also takes a look at what data compression related computations have successfully been ported to GPGPU platforms.

The rest of the paper is split into three sections. The next section~\ref{sec:gpgpu} describes the GPGPU platforms a bit closer, relating their characteristics to the needs of data compression. Section~\ref{sec:algorithms} shows some successful applications of GPGPU on compression algorithms, and describes what parts of a bzip2 implementations could be accelerated by GPGPUs. Finally in section~\ref{sec:discussion} we discuss the current trends in the field of GPGPU compression.

\section{GPGPU}
\label{sec:gpgpu}
Current leading edge GPU chips provide way higher computations per second compared to similarly priced leading edge CPU. The computational power of the GPUs is distributed over hundreds of computational cores, all working in parallel. To simplify synchronization of their concurrent computations, the GPGPU platforms run code in clusters of 32 threads called \emph{warps}. Each of the threads in one warp executes the same instruction, on different data, at every given time. For fast performance, the data the processors operate on should be coalesced in continuous and aligned blocks in memory, to enable a single memory transfer for all data. Multiple warps are run in blocks of hundred of threads, which combine into a grid consisting of all executing threads. Until recently all threads would always run the same program, but newer architectures have enabled simultaneous execution of multiple programs.

This programming model fits nicely in image processing where each computing core can process a single pixel, comparing it to nearby pixels in temporal or spatial dimensions, producing reductions or sensory compensations of them. On the other hand serial computations are slow on GPUs, as are many algorithms efficiently parallelizable on symmetric multiprocessing CPU systems. The limitations created by the tightly synchronized threads, and the memory access patterns restrict what is feasible to do on a GPGPU platform.

The split system with the CPUs on the motherboard, and the GPUs as separate devices also lead to some obstacles. Firstly all data data that is to be processed on GPGPU needs to be copied over the PCIe bus. Although the PCIe 2.0 bus is fast (8 GBPS) in comparison to many other transportation mediums, it does fall short of the parallel memory bandwidth on GPU cards ($>$200GBPS), which might lead to limits when wanting to do really fast stuff. Although the memory transfer from device (GPU) to host (CPU) can be performed in parallel with executed code, it always adds some extra latency to the overall runtime.

\subsection{GPGPU Data Compression}
The fact that lossless data compression is largely about finding redundancy within a set of data, leads to compression algorithms that process the data in one or many passes, comparing incoming data with previously encountered data. This makes many of them highly serial in nature, and are usually challenging to parallelize. The most common way of parallelizing compression algorithms is thus to split the data into smaller blocks and execute the normal serial algorithm on each one. As a downside, this separates the redundancies, giving slightly poorer compression ratio. There are multiple bzip2 implementations doing this  with efficient scaling on multiple CPU cores~\cite{gilchrist2009pbzip2}. This simple trick means that little research has been done on more efficiently parallelizable algorithms before GPGPU came along.

The fact that GPGPUs have hundreds of processors and a non-general memory accessing scheme, means that the method of splitting the data into smaller chunks isn't always feasible~\cite{wu2009cs315a}. On the other hand if one aims at very trivial and simple compression by processing single fields or integers into more compact form, using for example null suppression or Elias delta codes, one can achieve amazing parallel performance~\cite{o2011floating}.

The real deal of GPGPU compression is though to be found form exploiting more general computational methods available for GPGPU platforms. Firstly, being such a widely used general tool, sorting has several efficient parallel implementations, such as merge sort executing $O(n \log n)$ steps, in $O(\log n)$ time for a input of size $n$. On modern hardware it gives about a tenfold speedup compared to CPU implementations~\cite{satish2009designing}.

Perhaps more interesting is the parallel implementation of counting prefix sums. This algorithm computes intermediate values in a tree-like fashion producing an output with $O(n)$ steps, in $O(\log n)$ time and being around $\sim$20x faster than CPU implementations on modern hardware~\cite{nguyen2007gpu}. This has interesting utilizations in many GPGPU algorithms such as sorting~\cite{leischner2009gpu} and lexicographic naming~\cite{sun2009parallel}. In the next chapter where we will look into some parallel implementations of classic compression methods, where we can see more uses for the prefix sum.

\section{Current compression algorithms}
\label{sec:algorithms}
This chapter looks at some widely used compression algorithms, also used in bzip2, to see what parts of them have been successfully parallelized on GPGPU platforms. Notice beforehand that although the algorithms in this chapter are successful at compressing the incoming data, a efficient GPGPU decompressor might not always be available.

\subsection{Burrows-Wheeler Transform (BWT)}
The bzip2 compression stack starts with BWT~\cite{burrows1994block} to reorder the characters in the data to a more compressible form. Since the transform is simply executed by sorting the rotations of the input data, we can use any efficient sorting algorithm available for GPGPU. As stated above, there are many available that would make the encoding possible in parallel giving reasonable speedups. Although no research papers seem to have been published on the subject yet, preliminary figures in the community give claims of reasonable speedup~\cite{waveaccess2011cuda,inikep2011cuda}.

Unfortunately the inverse Burrows-Wheeler transform isn't as simple to parallelize. One possibility for parallelization would be to utilize the property of the transform that the inverse can be started from any point on the output. Given this we can on each parallel processor start at a point in the transformed string and continue until we hit a character from which decoding has been started by another processor. Problems would possibly arise from poor GPU performance of the very random memory accessing caused by the scattering of characters throughout the string.
\newpage

\subsection{Run-Length Encoding (RLE)}
\begin{figure}[t]
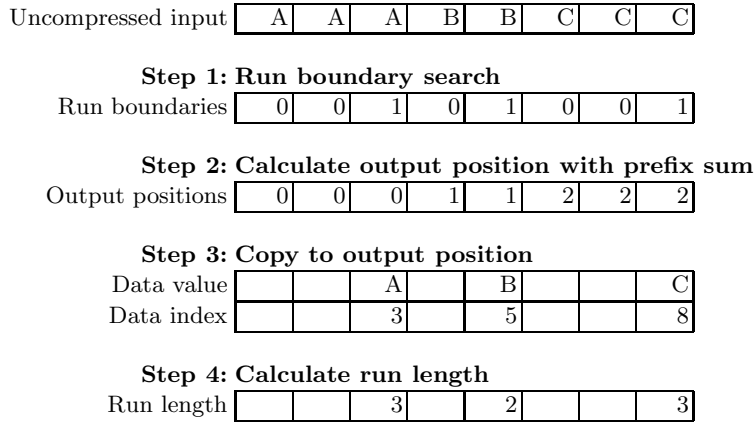

	\centering
	\begin{tabular}{rl}
	    Uncompressed input & 
		\begin{tabularx}{0.5\textwidth}{|R|R|R|R|R|R|R|R|}
			\hline
		    A & A & A & B & B & C & C & C \\
			\hline
		\end{tabularx} \\
		\\
	    \textbf{Step 1:} & \textbf{Run boundary search} \\
	    Run boundaries & 
		\begin{tabularx}{0.5\textwidth}{|R|R|R|R|R|R|R|R|}
			\hline
		    0 & 0 & 1 & 0 & 1 & 0 & 0 & 1 \\
			\hline
		\end{tabularx} \\
        \\
	    \textbf{Step 2:} & \textbf{Calculate output position with prefix sum}\\
	    Output positions & 
		\begin{tabularx}{0.5\textwidth}{|R|R|R|R|R|R|R|R|}
			\hline
		    0 & 0 & 0 & 1 & 1 & 2 & 2 & 2 \\
			\hline
		\end{tabularx} \\
	    \\
	    \textbf{Step 3:} & \textbf{Copy to output position} \\
	    Data value & 
		\begin{tabularx}{0.5\textwidth}{|R|R|R|R|R|R|R|R|}
			\hline
		      &   & A &   & B &   &   & C \\
		\end{tabularx} \\
	    Data index & 
		\begin{tabularx}{0.5\textwidth}{|R|R|R|R|R|R|R|R|}
			\hline
		      &   & 3 &   & 5 &   &   & 8 \\
			\hline
		\end{tabularx} \\
	    \\
	    \textbf{Step 4:} & \textbf{Calculate run length} \\
	    Run length & 
		\begin{tabularx}{0.5\textwidth}{|R|R|R|R|R|R|R|R|}
			\hline
		      &   & 3 &   & 2 &   &   & 3 \\
			\hline
		\end{tabularx} \\
	\end{tabular}
	\caption{Run-length encoding. Empty cells are not part of the array, but are added to retain the straight flow of values. Thus the last three arrays are of size 3.}
	\label{fig:rle_enc}
\end{figure}

Moving into the next step of the bzip2 compression, we encounter RLE, which compresses long runs of same characters into a single character along with a integer specifying the amount of times the character was repeated. Since RLE only focuses on very local redundancies between the characters, it is easier to design parallel algorithms for it. One such algorithm has been researched by W. Fang et al.~\cite{fang2010database}, and similar work has been done by A. Balevic~\cite{balevicfine}.

The algorithm described by Fang utilizes as a central part the parallel implementation for counting prefix sums, mentioned above. The encoding works in four parallel steps, as can be seen in figure~\ref{fig:rle_enc}. In the first step the boundaries between the runs are identified by having each processor compare its designated character with the neighboring one, outputting 1s in an auxiliary array to mark the boundaries. In the second step the prefix sum of the auxiliary array is computed to get the output positions for the characters and run lengths. In step 3 the algorithm simply copies the index of the boundary in the original data along with the character into the output array. The final step, executed on the now smaller output array, computes the lengths of the runs by comparing the differences of the indices.

\begin{figure}[th]
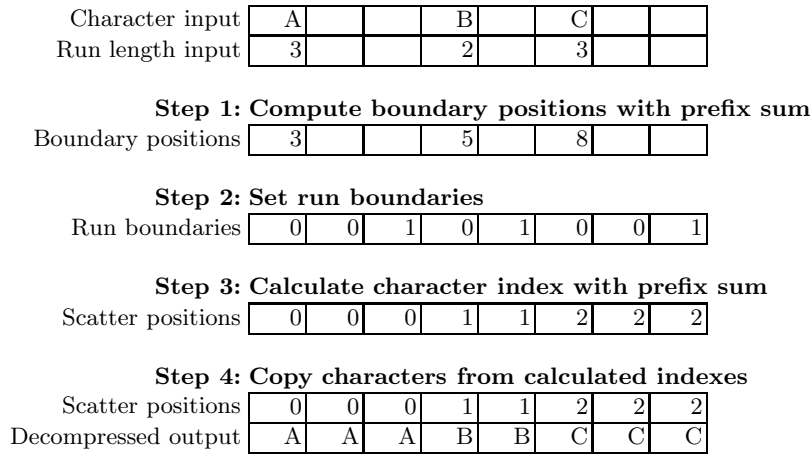

	\centering
	\begin{tabular}{rl}
	    Character input & 
		\begin{tabularx}{0.5\textwidth}{|R|R|R|R|R|R|R|R|}
			\hline
		    A &   &   & B &   & C &   &   \\
		\end{tabularx} \\
	    Run length input & 
		\begin{tabularx}{0.5\textwidth}{|R|R|R|R|R|R|R|R|}
			\hline
		    3 &   &   & 2 &   & 3 &   &   \\
			\hline
		\end{tabularx} \\
        \\
	    \textbf{Step 1:} & \textbf{Compute boundary positions with prefix sum} \\
	    Boundary positions & 
		\begin{tabularx}{0.5\textwidth}{|R|R|R|R|R|R|R|R|}
			\hline
		    3  &   &   & 5 &   & 8 &   &   \\
			\hline
		\end{tabularx} \\
        \\
	    \textbf{Step 2:} & \textbf{Set run boundaries} \\
	    Run boundaries & 
		\begin{tabularx}{0.5\textwidth}{|R|R|R|R|R|R|R|R|}
			\hline
		    0 & 0 & 1 & 0 & 1 & 0 & 0 & 1 \\
			\hline
		\end{tabularx} \\
        \\
	    \textbf{Step 3:} & \textbf{Calculate character index with prefix sum} \\
	    Scatter positions & 
		\begin{tabularx}{0.5\textwidth}{|R|R|R|R|R|R|R|R|}
			\hline
		    0 & 0 & 0 & 1 & 1 & 2 & 2 & 2 \\
			\hline
		\end{tabularx} \\
        \\
	    \textbf{Step 4:} & \textbf{Copy characters from calculated indexes} \\
	    Scatter positions & 
		\begin{tabularx}{0.5\textwidth}{|R|R|R|R|R|R|R|R|}
			\hline
		    0 & 0 & 0 & 1 & 1 & 2 & 2 & 2 \\
		\end{tabularx} \\
	    Decompressed output & 
        \begin{tabularx}{0.5\textwidth}{|R|R|R|R|R|R|R|R|}
			\hline
		    A & A & A & B & B & C & C & C \\
			\hline
		\end{tabularx} \\
		
	\end{tabular}
	\caption{Run-length decoding}
	\label{fig:rle_dec}
\end{figure}

RLE is one of the few GPGPU compression algorithms for which a efficient decoding has been found and implemented. Working similarly as the encoding in reverse, figure~\ref{fig:rle_dec} shows the four steps of the algorithm. First the indices of the boundaries are calculated from the run lengths using the prefix sum algorithm. Secondly 1s are written into an auxiliary array at the boundaries in the previous result. In step 3 the prefix sum is run on the auxiliary array to compute an array containing indices of the compressed input. In the last step characters are copied to the output array according to the indices computed in the previous step.

\subsection{Move-to-front}
Although no published research could be found on MTF encoding, there seems to be a possibility for trivial parallelization. By parallelizing over the stack of recently used symbols to find the index of the read character, we could reduce the time of looking up the index of the character. In practice each processor would check if the character at the processors index matches the input character and output this index to some shared register. Since each character occurs in the stack only once, no concurrency race conditions will occur for the write. Updating the state of the stack can also be similarly parallelizable.

This would lead to constant time lookup of the index, provided that the GPU has equal amount or more processors than amount characters in the alphabet. Although every core in the GPU is easily utilized, one must take note that this implementation would be executing more redundant computations compared to a serial implementation which would stop after finding the first, and only, occurrence.

Speeding up decoding on GPGPU platforms might be more challenging since the character lookup is already constant time on serial implementations, and starting decoding from multiple places is difficult since the state of the stack is not known at the other places.

\subsection{Variable-Length Encoding (VLE)}
\begin{figure}[th]
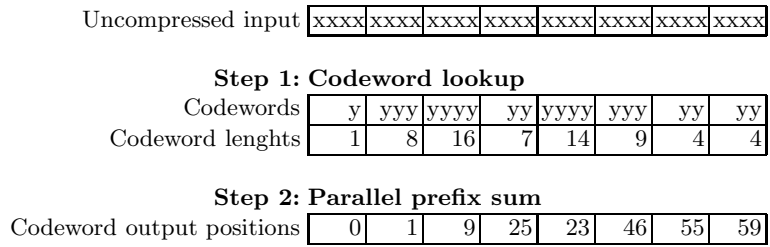

	\centering
	\begin{tabular}{rl}
	    Uncompressed input & 
		\begin{tabularx}{0.5\textwidth}{|R|R|R|R|R|R|R|R|}
			\hline
		    xxxx & xxxx & xxxx & xxxx & xxxx & xxxx & xxxx & xxxx \\
			\hline
		\end{tabularx} \\
		\\
	    \textbf{Step 1:} & \textbf{Codeword lookup} \\
	    Codewords & 
		\begin{tabularx}{0.5\textwidth}{|R|R|R|R|R|R|R|R|}
			\hline
		    y & yyy & yyyy & yy & yyyy & yyy & yy & yy \\
		\end{tabularx} \\
	    Codeword lenghts & 
		\begin{tabularx}{0.5\textwidth}{|R|R|R|R|R|R|R|R|}
			\hline
		    1 & 8 & 16 & 7 & 14 & 9 & 4 & 4 \\
			\hline
		\end{tabularx} \\
        \\
	    \textbf{Step 2:} & \textbf{Parallel prefix sum} \\
		Codeword output positions &
		\begin{tabularx}{0.5\textwidth}{|R|R|R|R|R|R|R|R|}
			\hline
		    0 & 1 & 9 & 25 & 23 & 46 & 55 & 59 \\
			\hline
		\end{tabularx}
	\end{tabular}
	\caption{Variable length encoding. xxxx represents one generic uncompressed character, and yyy represents it's generic codeword.}
	\label{fig:vle_enc}
\end{figure}

The last step of bzip2 covered in this paper is the Huffman coding~\cite{huffman1952method}. Although no GPGPU algorithms are at the time available for constructing the actual Huffman tree, there exists at least one algorithm by A. Balevic~\cite{balevic2010parallel}, focusing on the substitution from uncompressed data to variable-length encoding. Although the substitution seems like a trivial task for a serial algorithm, this isn't the case when we want to parallelize. The problem arises from the fact that due to variable length codewords, we don't know where into the output array the converted codes should be copied.

The encoding algorithm works as in figure~\ref{fig:vle_enc}. First the input characters are converted to codewords in parallel, also storing the corresponding codeword length into an auxiliary array. Secondly the prefix sum is calculated from the auxiliary array to get the output positions of the codewords. Finally the codewords are copied into the calculated indexes.

Here again, decompression is harder. This is due to the fact that the decoder doesn't know where one codeword ends and another begins before it has decoded the whole prior input. Some work on parallelizing variable-length codes using error-resilience methods has been done~\cite[Ch. 3.10]{biskup2008error}, and could possibly be used on GPGPU platforms.

\subsection{Other compression usages}
In addition to general lossless data compression focusing on large scale redundancies as in the case of bzip2, successful work has also been done in other types of GPGPU data compression applications. Firstly a team at the Texas state university has implemented a high-speed floating point compression algorithm called GFC~\cite{o2011floating} that touts 75GB/s compression speeds for usage in scientific computation where massive amounts of data are generated and need to be processed in real time. The compression methods used are quite simple, mostly focusing on only storing deltas between floating point values and removing unnecessary zeroes form the result. Unfortunately, the practical processing speed of the compression is limited by the PCIe bus to 8GB/s.

Other work has been done in database data compression~\cite{fang2010database}, where multiple different light-weight compression methods are combined to speed up GPU co-processing of database queries. Compressing the data makes it faster to move the data over the PCIe bus, and faster to read from disk. The compression techniques focus on compression of single database fields such as integers, using null-suppression, dictionaries, RLE and more.

\section{Discussion}
\label{sec:discussion}
The compression algorithms described above compose the central parts of bzip2. Given proper implementations of each, it shouldn't be hard to combine them togther for a bzip2 compatible compressing tool that could deliver high performance on modern GPUs. The efficiency of the MTF algorithm is debatable. But the benefit of having all code running on the GPU, removing unnecessary data transfer over the PCIe bus, might still make the overall compression speed faster.

As for the codeword tables for the VLE, it seems unlikely that efficient Huffman-tree GPGPU algorithms will be possible. But this doesn't mean that pre-calculated ones couldn't be used, combined with some data analysis for choosing a fitting one for the current input data.

Although most parallel compression methods depend on splitting the compressible data into small chunks, cutting the dependancies between the chunks, we have here taken a look at some more sophisticated attempts at parallel compression on GPGPU platforms. GPGPU compression focusing on local changes can give very high compression bandwidths, while at the same time more wide scale algorithms are using general computational primitives for efficient parallelizable implementations.

Future work will most likely be affected by the different general algorithms discovered for GPGPU platforms, and made easily available by libraries such as Thrust~\cite{thrust}.
\bibliographystyle{splncs_srt}
\bibliography{gpgpuc}

\end{document}